\documentclass[twoside,english,english,aps,prd,amsmath,amssymb,superscriptaddress]{revtex4}
\usepackage[T1]{fontenc}
\usepackage[latin9]{inputenc}
\usepackage{mathrsfs}
\usepackage{amsmath}
\usepackage{amsthm}
\usepackage{amssymb}
\usepackage{stmaryrd}

\makeatletter
\@ifundefined{textcolor}{}
{%
 \definecolor{BLACK}{gray}{0}
 \definecolor{WHITE}{gray}{1}
 \definecolor{RED}{rgb}{1,0,0}
 \definecolor{GREEN}{rgb}{0,1,0}
 \definecolor{BLUE}{rgb}{0,0,1}
 \definecolor{CYAN}{cmyk}{1,0,0,0}
 \definecolor{MAGENTA}{cmyk}{0,1,0,0}
 \definecolor{YELLOW}{cmyk}{0,0,1,0}
}
\numberwithin{equation}{section}
\numberwithin{figure}{section}

\makeatother

\usepackage{babel}
\begin{document}
\title{Ricci Flow Approach to The Cosmological Constant Problem}
\author{M.J.Luo}
\address{Department of Physics, Jiangsu University, Zhenjiang 212013, People's
Republic of China}
\email{mjluo@ujs.edu.cn}

\begin{abstract}
In order to resolve the cosmological constant problem, the notion
of reference frame is re-examined at the quantum level. By using a
quantum non-linear sigma model (Q-NLSM), a theory of quantum spacetime
reference frame (QSRF) is proposed. The underlying mathematical structure
is a new geometry endowed with intrinsic 2nd central moment (variance)
or even higher moments of its coordinates, which generalizes the classical
Riemannian geometry based on only 1st moment (mean) of its coordinates.
The 2nd central moment of the coordinates directly modifies the quadratic
form distance which is the foundation of the Riemannian geometry.
At semi-classical level, the 2nd central moment introduces a flow
which continuously deforms the Riemannian geometry driven by its classical
Ricci curvature, which is known as the Ricci flow. A generalized equivalence
principle of quantum version is also proposed to interpret the new
geometry endowed with at least 2nd moment. As a consequence, the spacetime
is stabilized against quantum fluctuation, and the cosmological constant
problem is resolved within the framework. With an isotropic positive
curvature initial condition, the long flow time solution of the Ricci
flow exists, the accelerating expansion universe at cosmic scale is
an observable effect of the spacetime deformation of the normalized
Ricci flow. A deceleration parameter -0.67 consistent with measurement
is obtained by using the reduced volume method introduced by Perelman.
Effective theory of gravity within the framework is also discussed.
\end{abstract}
\maketitle

\section{Introduction}

The incompatibility between Quantum Mechanics (QM) and General Relativity
(GR) is not only reflected at the technical level that GR is not renormalizable
in ordinary sense, but also at the phenomenology level that GR and
the spacetime is fundamentally unstable at the quantum level. If an
effective quantum field theory arises infinity in its calculation,
it does not necessarily imply a complete disaster. It may call one's
attention to treat the parameters of the theory more seriously that
one should discriminate which part of the parameters are physical
and which part are not. However, the incompatibility between QM and
GR is not just simply the case: if QM is seriously taken into account,
the classical spacetime depicted by GR will rapidly collapse and even
impossible to exist. More precisely, this severe difficulty concerning
the instability of spacetime under quantum fluctuation names the Cosmological
Constant (CC) problem. 

If someone may think that the CC problem is a non-essential side issue
of physics, we consider it is a crisis of fundamental physics \cite{weinberg1989cosmological}.
The CC arises as a severe problem is not at the classical level but
at the quantum level, because anything including the spacetime is
inescapable quantum fluctuating, and the well-tested Equivalence Principle
(EP) claims that any quantum fluctuation contributing to the energy
density of the vacuum couples to gravity and behaves like a CC. A
standard calculation only concerning the zero-point vacuum oscillating
modes up to the Planck scale $\Lambda_{pl}$, below which the calculations
are trustable, gives the energy density $\sim\Lambda_{pl}^{4}$ which
shows that the vacuum energy densities of quantum fluctuations and
CC (if we trust the well-tested EP) should be too large to make the
spacetime stable and permanently exist. However, the observation from
the accelerating expansion of the universe \cite{Perlmutter:1998np,Riess:1998cb,Ade:2015xua}
shows that the CC is much smaller compared with the prediction. Why
so large amount of vacuum energy densities do not gravitate? And if
they could be canceled by certain unknown mechanism (e.g. supersymmetry),
why they just leave a small remnant to the gravitational effect (i.e.
accelerating expansion)? The problem leads to a severe fine-tuning
to make the spacetime the way it is under the quantum fluctuation,
just like to fine-tune a sharpened pencil standing on a table against
perturbation. 

There are many attempts to solve the CC problem, one can find good
review articles, e.g. \cite{2008GReGr..40..607B,amendola2010dark,miao2011dark,2012CRPhy..13..566M,sola2013cosmological},
and references therein. The attempts cover from the phenomenological
models, tuning mechanisms, modified gravity, to even anthropic principle.
The fundamental incompatibility is sometimes ignored and evaded. 

At the fundamental level, one may puzzle that if the quantum fluctuation
is real whether the EP is wrong at the quantum level? The fact is
that it is well-known that the electron vacuum energy coming from
the vacuum polarization measured by the Lamb's shift does gravitate
normally as the EP claims \cite{Polchinski:2006gy,2009PhLB..679..433M}.
There is no any evidence that the energy coming from classical and
quantum are physically different, the EP is well-tested at very precise
level. Essentially speaking, physicists are caught in a dilemma that
both quantum fluctuations and the EP are so real and precisely tested
in each field, why they give rise to an obvious wrong prediction.

In the paper we start with the assumptions that the validity of the
EP is retained and generalized even to the quantum level, and the
realness of the quantum fluctuations are also admitted. The approach
to resolve the dilemma proposed in the paper is twofold, on the one
hand the spacetime geometry is treated in a more quantum manner (Chapter
2) via the Ricci flow approach (Chapter 3), and on the other hand
the principle of QM is treated in a more relational manner \cite{1996IJTP...35.1637R}
(Chapter 4) via a framework of entanglement that a to-be-studied quantum
system relative to a Quantum Spacetime Reference Frame (QSRF) system
\cite{Luo2014The,Luo2015Dark,Luo:2015pca}. The Ricci flow describes
a continuous deformation of a Riemannian geometry from short distance
to long distance scale induced by the quantum fluctuation of the spacetime,
which makes the geometry of the universe more and more like an observed
accelerating expansion universe (Chapter 5), or equivalently, develops
an effective CC consistent with observation in the gravity theory. 

The Ricci flow was introduced in 1980s in mathematics by Hamilton
\cite{Hamilton1982Three,hamilton1986four,hamilton1988}. Hamilton
used it as a tool to gradually deform a manifolds into a more and
more ``nice'' manifolds whose topology is easily recognized, in
order to prove the Poincare's conjecture. The program was fully realized
owing to Perelman's breakthrough around 2003 \cite{perelman2002entropy,perelman2003ricci,perelman307245finite}
by introducing some monotonic functionals to successfully deal with
the singularities developed in 3-manifolds under the Ricci flow. The
Ricci flow approach (see reviews e.g. \cite{chow2004ricci,topping2006lectures,chow2006hamilton,morgan2007poincare,muller2006differential,chow2007ricci})
as a useful tool in mathematics may have important physical applications,
e.g. see \cite{gutperle2003spacetime,Bakas:2005kv,headrick2006ricci,samuel2007geometric,solodukhin2007entanglement,tseytlin2007sigma,woolgar2008some,carfora2010renormalization,headrick2010new,figueras2011ricci,ivancevic2011ricci},
including early attempt applying it to the cosmology as an averaging
approach to the spatial inhomogeneous \cite{Carfora1984Smoothing,carfora1988model,carfora1995renormalization,piotrkowska1995averaging,carfora2008ricci}.
The Ricci flow was also found in 1980s by Friedan as an approximate
RG flow of a 2d non-linear sigma model (2d-NLSM) \cite{friedan1980nonlinear,Friedan1980}.
The NLSM can also be used to generate an effective CC for extra dimensions
in Kaluza-Klein models \cite{1980Generalized}, the dynamics of spacetime
coordinates being used to generate extra contributions to the energy-momentum
has also been discussed in \cite{Giddings:2005id}. But to the best
of our knowledge, the physical meaning of Ricci flow and NLSM in physics
is not very clear in the community, and their connections to the CC
problem have also not been clearly clarified, the goal of the paper
is to show their deep relation.

The solution to the CC problem and the dilemma can be briefly stated
as follows. The notion of parameter spacetime coordinates that is
external and free from quantum fluctuation must be abandoned, so the
quantum zero-point fluctuation energies of vacuum be relative to the
parameter spacetime coordinates are completely unobservable and unphysical,
including the Casimir effect \cite{PhysRevD.72.021301}. When they
are relative to a more physical quantum spacetime reference system
(i.e. QSRF) which is also zero-point fluctuating quantum mechanically,
the vacuum energy density is obtained a correct value. And as the
EP claims, the vacuum energy densities universally coupled to gravity,
giving rise to the uniform accelerating expansion universe at cosmic
scale. The accelerating expansion universe at long distance scale
is an observable effect of the spacetime deformation via a normalized
Ricci flow.

\section{Spacetime with Intrinsic 2nd Moments: Non-linear Sigma Model}

In order to reconcile the incompatibility between QM and GR and hence
resolve the cosmological constant problem, a question at least is
how to consistently apply the principle of QM to the spacetime geometry,
avoiding the instability of spacetime against quantum fluctuations.
It is generally believed that we need a new framework of spacetime
geometry based on quantum rulers and clocks. In classical geometry,
the Riemannian geometry, the central concept is to measure the length
between two point coordinates. In the language of QM, we measure the
mean value or the 1st moment of a coordinate. However, we could imagine
that under quantum fluctuation, coordinates of the geometry smear
and hence higher moments of a coordinate in measurement naturally
appear, for instance, the 2nd central moment, the variance of the
coordinate. A crucial question is how to introduce the higher moments
to a geometry, making them well-behaved under both the principle of
QM and geometry. We find that the classical Riemannian geometry does
not explicitly contain the notion of (higher) moments in it. To our
knowledge, a new geometry with well-behaved higher moments has not
been developed yet. However, the classical Riemannian geometry is
not far from the new geometry, because it is a good approximation
at the level of 1st moment, the mean value of coordinates, if the
variance is not large enough. In the section, we suggest a generalization
of Riemannian geometry by considering higher moments on it.

To define the geometry of D-dimension with at least 2nd moment, we
construct a non-linear differentiable mapping $X$ from a local coordinate
patch $x\in\mathbb{R}^{d}$ to a manifolds $M^{D}$. The mapping in
physics can be realized by a kind of fields theory, the non-linear
sigma model (NLSM) \cite{gell1960axial,friedan1980nonlinear,Friedan1980,zinn2002quantum,ketov2013quantum,2015Non},
\begin{equation}
S_{X}=\frac{1}{2}\lambda\int d^{d}xg_{\mu\nu}\frac{\partial X^{\mu}}{\partial x^{a}}\frac{\partial X^{\nu}}{\partial x^{a}}.\label{eq:NLSM}
\end{equation}

A quantum fields theory is usually formulated in an inertial frame,
for instance, the base space. The base space of NLSM $x^{a}$, $a=0,1,...d-1$
can be interpreted as a flat Laboratory Wall Frame (LWF) (for detail
see \cite{Luo2014The,Luo:2015pca}). The walls and clock of a laboratory
can be used as starting references to orient, align and order the
beams of the scalar particles $X^{\mu}$ at high precision, in this
sense, these identical (quantum) scalar particles can be interpreted
as physical rulers and clock, or frame fields (at quantum level).
For example, the identical scalar particles oriented as $X_{1,2,3}=X,Y,Z$
can be aligned with a reference to the $x,y,z$-directions of the
walls of the laboratory, respectively. One could visualize them as
local quantum vibrations or oscillations placed on the lattice of
$x,y,z$, these identical particles can be seen as rulers since distances
could be measured by counting their phase changes of the local vibrations
if events trigger the counting. In this sense they play the roles
of state-triggers labeling where the event happens. For the similar
consideration, a scalar particle $X_{0}=T$ on the lattice can be
used to play the role of a small pendulum clock labeling the causal
order of the events, i.e. when the event happens. A practical example
for the scalar particles model of spacetime is the multi-wire proportional
chamber: the beams of the scalar particles $X^{\mu}$ used in the
model are practically electrons in the array of multi-wire, which
signal the coordinates of an event by an impulse at the output. Certainly,
the output electron signal is inescapably quantum fluctuating.

In the context of quantum fields theory, the base-space-dimension
$d$ of the theory is not an observable, $d$ as an input parameter
of the theory may run as the theory is renormalized, so we do not
fix $d$ from the beginning. While according to the required LWF interpretation
of $d$ at the laboratory scale, we could set a renormalization point
at which scale giving $d\approx4$ in common sense (the meaning of
''$\approx$'' will be explained later). However, it is also possible
that at other scale $d$ may not necessarily be $\approx4$, it may
become fractal $d<4$ and even run to $d=2$ at UV so that the model
seems effectively power counting renormalizable. Literature \cite{Luo:2015pca}
suggests a hand waving reasoning to the dimensional running, but the
exact mechanism of dimensional running is still unclear.

The mapping $X^{\mu}$ ($\mu=0,1,...D-1$) in NLSM, as the scalar
frame fields or the coordinates of the manifolds $M^{D}$ here, can
be interpreted as the spacetime reference frame \cite{Luo:2015pca},
so we fix $D=4$. The metric $g_{\mu\nu}$ of the manifolds is defined
by comparisons between the distant frame fields $X^{\mu}$ and the
native clock in the laboratory, that is, $g_{\mu\nu}\equiv\frac{\partial X_{\mu}}{\partial x_{a}}\frac{\partial X_{\nu}}{\partial x_{a}}$.
Although we could align the beams of particles $X^{\mu}$ in LWF with
reference to the walls of the laboratory, when the geometric measurements
are extrapolated to a larger scale much beyond the laboratory wall,
the beams of the particles $X^{\mu}$ can only locally self-reference
by the metric $g_{\mu\nu}$ at a distant point $X$. Since quantum
fluctuations are inescapable in $X^{\mu}$, the coordinates of the
spacetime geometry at distant must subject to some level of quantum
fuzziness, making the distant clocks in principle can not be synchronized
precisely with the native clock. At quantum level, it means that beside
the mean value $\langle X^{\mu}\rangle$ of the coordinates, it is
inescapably to consider the 2nd (central) moment $\langle(\delta X^{\mu})^{2}\rangle=\langle(X^{\mu})^{2}\rangle-\langle X^{\mu}\rangle^{2}$
measuring the quantum variance or the quantum fuzziness of the coordinates.
More generally, we must consider $\langle\delta X^{\mu}\delta X^{\nu}\rangle$
measuring the quantum covariance. In the situation, the expectation
value of metric $\langle g_{\mu\nu}\rangle$ must be deformed by the
2nd (central) moment as the definition indicates and hence curved
in general. 

In the paper, we often ignore the phrase ``central'', when mentioning
``moment'', we always mean the ``central moment'' for short, and
the bracket $\langle.\rangle$ always mean quantum expectation value.
Similar to the 2nd moment, the 3rd moment ``skewness'' $\langle(\delta X^{\mu})^{3}\rangle$
or more general $\langle\delta X^{\mu}\delta X^{\nu}\delta X^{\rho}\rangle$
describes the asymmetry of the quantum fuzziness, the 4th moment ``kurtosis''
and other higher moments may also exist. If the 2nd moment is most
important, it is Gaussian, and the moments higher than 2nd order we
call them non-Gaussian fluctuations. 

Note that the classical solution of the NLSM is $\langle X^{\mu}\rangle=\langle e_{a}^{\mu}\rangle x^{a}$,
$e_{a}^{\mu}$ the vielbein, so in the sense of classical Riemannian
geometry, the mapping $X^{\mu}$ is just a classical coordinate transformation
of $x^{a}$. However, because of the quantum fluctuations of the coordinates
$X^{\mu}$ and the existence of higher moments of coordinates, the
quantized NLSM (Q-NLSM) in fact defines a new geometry with intrinsic
higher moments beyond the classical Riemannian geometry, which is
not clearly developed before. In the sense, the NLSM describing a
classical Riemannian geometry is nothing but a 1st moment approximation
to the new geometry. The quantum fluctuation of the Q-NLSM in fact
introduces some extra (quantum) structures to the classical Riemannian
geometry, which leads to new phenomenon of spacetime and gravity such
as coarse graining process of the spacetime geometry. 

There are evidences that this kind of Q-NLSM (\ref{eq:NLSM}) may
be a good laboratory to study effects of quantum gravity: (1) There
are deep analogies between Einstein's theory of gravity and the NLSM
\cite{percacci2009asymptotic}; (2) it may has a non-trivial or non-Gaussian
UV fixed point (an interacting or non-flat fixed point) at non-perturbative
level so that it is asymptotically safe and renormalizable \cite{codello2009fixed,wellegehausen2014asymptotic}.
However, whether the model (\ref{eq:NLSM}) is rigorously renormalizable
and rigorously a quantum theory of gravity is beyond the scope of
the paper and not our purpose, frankly speaking, the proof is not
an easy task yet. Since the renormalizability of the NLSM depends
on $d$, but if we want to give a proper (QSRF) interpretation to
the NLSM, certain power counting non-renormalizable dimension must
be taken at certain scale or even a running dimension is possible,
but the mechanism of dimensional running is not exactly known, so
the purpose of the paper is to use a more or less $d$-insensitive
and well-developed approach to see whether the flow (driven by the
2nd moment so defined) of the geometry can give an observed accelerating
expansion spacetime, we see this kind of approximate approach have
already developed, it is the Ricci flow.

\section{Geometric Flow driven by 2nd Moment: The Ricci Flow}

Here we consider the effects of the lowest 2nd (central) moment, the
variance, in the geometry. The classical Riemannian geometry is a
kind of manifolds based on a metric of quadratic form, i.e. the length
measured between two points coordinates. The mean value of the coordinates
is not affected by the 2nd moment. However, because in classical Riemannian
geometry, the distance is a quadratic form of the coordinates, the
2nd moment now gives an extra contribution (i.e. fluctuation $\delta X$)
to the distance. The distance as a quadratic form is generalized to
\begin{equation}
\langle\Delta X^{\mu}\Delta X_{\mu}\rangle=\langle\Delta X^{\mu}\rangle\langle\Delta X_{\mu}\rangle+\langle\delta X^{\mu}\delta X_{\mu}\rangle,
\end{equation}
where we always denote $\Delta X$ the classical displacement of $X$
and $\delta X$ the fluctuation relating to the 2nd moment contribution. 

Since a local distance element in Riemanian geometry is given by a
local metric tensor at a point, so equivalently, it is convenient
to think of the location point $X$ being fixed, and we interpret
the 2nd moment of coordinates affects only the metric tensor $g_{\mu\nu}$
at the location point $X$, then the expectation value of $g_{\mu\nu}$
including the 2nd moment can be given by,
\begin{equation}
\langle g_{\mu\nu}(X)\rangle=\left\langle \frac{\partial X_{\mu}}{\partial x_{a}}\frac{\partial X_{\nu}}{\partial x_{a}}\right\rangle =\frac{\partial\langle X_{\mu}\rangle}{\partial x_{a}}\frac{\partial\langle X_{\nu}\rangle}{\partial x_{a}}+\frac{1}{(\Delta x)^{2}}\langle\delta X_{\mu}\delta X_{\nu}\rangle=g_{\mu\nu}^{(1)}(X)+\delta g_{\mu\nu}^{(2)}(X).\label{eq:g+dg}
\end{equation}
The first term $g_{\mu\nu}^{(1)}(X)=\frac{\partial\langle X_{\mu}\rangle}{\partial x_{a}}\frac{\partial\langle X_{\nu}\rangle}{\partial x_{a}}=\langle e_{\mu}^{a}\rangle\langle e_{\nu}^{a}\rangle$
is the 1st moment contribution from the classical Riemannian geometry.
The second term $\delta g_{\mu\nu}^{(2)}$ proportional to the 2nd
moment $\langle\delta X_{\mu}\delta X_{\nu}\rangle$ is a quantum
correction or deformation to the classical 1st moment metric, and
$\delta g_{\mu\nu}^{(2)}$ also depends on a cutoff length scale of
the LWF ($(\Delta x)^{2}=\eta_{ab}\Delta x_{a}\Delta x_{b}$) relating
to a minimum resolution in measuring the metric $\langle g_{\mu\nu}(X)\rangle$
in the observer's laboratory.

The 2nd moment of the geometry as an extra quantum structure introduced
by the Q-NLSM can be determined by the 2-point correlation function
in the Q-NLSM. We derive the flow equation of $\delta g_{\mu\nu}^{(2)}$
at the laboratory or classical scale by setting $d\approx4$ in (\ref{eq:NLSM}).
In this setting, the base space is a 4-dimensional LWF, and its semi-classical
approximation of the theory (shown later) can reproduce the classical
theory. 

If we denote $R^{(1)}$ as the scalar curvature given by $g_{\mu\nu}^{(1)}$
at the point $X$, $\delta k^{2}$ the cutoff energy scale playing
the role of the inverse of the cutoff length $(\Delta x)^{2}$, and
$\lambda$ the prefactor of the NLSM, in the condition $R^{(1)}(X)\delta k^{2}\ll\lambda$,
at lowest order perturbative calculation we have \cite{codello2009fixed}
\begin{equation}
\delta g_{\mu\nu}^{(2)}=\frac{R_{\mu\nu}^{(1)}}{32\pi^{2}\lambda}\delta k^{2}.\label{eq:ricci-flow-k}
\end{equation}
In this equation, we assume that the 2nd moment contribution $\delta g_{\mu\nu}^{(2)}$
is smaller compared with the 1st moment $g_{\mu\nu}^{(1)}$ and it
is larger than other higher order moments, so it is a perturbative
and Gaussian approximation to the RG flow of NLSM. We will later see
that the physical interpretation of $\lambda$ is the critical density
of the universe $\lambda=\frac{3H_{0}^{2}}{8\pi G}$, where $H_{0}$
is the currently observed Hubble's constant and $G$ the Newton's
constant. Therefore, when we discuss the problem of cosmology, the
scalar curvature is approximately $R^{(1)}\sim O(H_{0}^{2})$ and
$\delta k^{2}\ll1/G\sim O(M_{p}^{2})$, so the perturbative condition
$R^{(1)}(X)\delta k^{2}\ll\lambda$ is safe. It is worth mentioning
that $\lambda$ as the unique parameter of the theory is a combination
of the Hubble's constant and Newton's constant, which differs from
the traditional gravity theory which only has the Newton's constant.
The Hubble's constant here comes into the fundamental theory, and
plays an important role in giving a characteristic scale of the universe.
So it will lead to new phenomenon lower than the characteristic energy
scale. From the equation (\ref{eq:ricci-flow-k}), we see that the
2nd moment correction to the metric is proportional to the Ricci curvature,
so it is reasonable to think that the 2nd moment can also give acceleration
or gravitational effects relating to the Ricci curvature as the EP
claims, if the EP is valid at quantum level as well, that is the simple
evidence why the 2nd moment can leads to an accelerating universe
at cosmic scale.

We conclude that in the approximation, the metric tensor with 2nd
moment seems like a classical metric with a deformation driven by
its classical Ricci curvature. The deformation introduces a flow of
metric tensor in Riemannian geometry, the (\ref{eq:ricci-flow-k})
is nothing but the standard Ricci flow equation
\begin{equation}
\frac{\partial g_{\mu\nu}}{\partial t}=-2R_{\mu\nu}^{(1)},\label{eq:ricci-flow-t}
\end{equation}
with 
\begin{equation}
t=-\frac{1}{64\pi^{2}\lambda}k^{2}.\label{eq:t}
\end{equation}
Because the 2nd moment only modifies the local quadratic form of length
which is not important for the topology of the geometry, so the deformation
by the Ricci flow does not change the topology, the flow preserves
it.

Averaging procedures in a non-linear gravity system are highly non-trivial
\cite{Zalaletdinov:2008ts,Paranjape:2009zu} and fundamentally an
unclear issue yet. The Ricci flow is in essential a process of averaging
or coarse graining of the non-linear gravity system. The Ricci flow
equation is a non-linear generalization of the heat flow. The $t$
parameter of the flow corresponds to the energy scale $k$, when the
flow starting from $t=-\infty$ flows forwardly to $t=0$, it is equivalent
to that the geometry starting from high energy scale $k\rightarrow\infty$
(short distance UV scale) flows forwardly to low energy $k\rightarrow0$
(long distance IR scale). At a cut off distance scale $1/k$, the
2nd moment being shorter than $1/k$ are averaged out and gives correction
to the quadratic form of length longer than the scale. As the geometry
deforms and flow to long distance, it losses its information of shorter
distance, so the flow is non-reversible. Because here the behavior
of the fluctuation of the manifolds is highly non-linear, the average
process is much complex than a conventional one, quantitatively it
describes by the non-linear Ricci flow equation.

The Ricci flow is a diffusion-reaction-like and non-linear equation,
in which the diffusion and reaction compete each other. For certain
initial condition of isotropic and positive curvature, as it flows
to long distance scale, the diffusion always win out, then the local
quantum fluctuations and related moments of the geometry are gradually
averaged out as the heat equation uniformlizes the temperature distribution,
the geometry with 2nd moments gradually deforms to a classical Riemannian
geometry with a uniform curvature. The phenomenon indicates that the
2nd moment and Gaussian fluctuation are irrelevant, the spacetime
stabilizes against the quantum fluctuation and becomes more and more
classical in the IR. The latter of the paper will show that it corresponds
to the observation of the accelerating expansion universe at cosmic
distance. For some anisotropic initial conditions, when the reaction
term wins at the place where local curvature is larger enough than
other places, the flow equation in general develops local singularities,
near which the Riemannian geometry becomes singular and no longer
can be used to model the geometry, at the time, the Ricci flow as
a Gaussian approximation of the geometry with 2nd moment fails. 

An important observation is that the Ricci flow equation is formulated
by using $t$ but $k$, so the flow does not explicitly care about
the base-space-dimension $d$ of NLSM which is hidden in a specified
form of $t$, e.g. $t=-\frac{C_{d}}{2\lambda}k^{d-2}$, $C_{d}^{-1}=(4\pi)^{d/2}\Gamma(\frac{d}{2}+1)$
for general $d$. So in this sense the Ricci flow can be applied for
general $d$ only with different allowed ranges of $t$. Strictly
speaking, it can be applied for all $d<D=4$, since the mapping $\mathbb{R}^{d}\rightarrow M^{D}$
of the NLSM are all trivial. As our particular interest in this paper,
if we consider the target space $M^{D}$ a homogeneous and isotropic
D-sphere, the homotopy group $\pi_{d}(S^{D=4})=0$ for all $d<D=4$,
so all the models with $d<4$ behave similar in the Ricci flow. The
case $d=D=4$ is a little tricky, since $\pi_{4}(S^{D=4})=\mathbb{Z}$
are not trivial, the mapping may meet intrinsic topological obstacle
and become singular, so the flow may not be convergent, which means
non-renormalizable. Thus we have to assume $d$ can be very close
to 4 but not precisely, denoted as $d\approx4$ always meaning $d=4-\epsilon$
($\epsilon$ a very small positive number) throughout the paper, meanwhile
$d$ is so close to 4 that the conventional topological arguments
based on $d=4$ are also applicable. In fact if the flow corresponding
to $d=4$ is at high energy region, far from the low energy topological
phase transition induced by singularity of the mapping, its high energy
behavior can be approximately seen similar with the cases $d<4$.
This property of the Ricci flow allows us to make some $d$-insensitive
observations to the Q-NLSM with certain $d$ which is required a proper
(QSRF) interpretation at the laboratory scale, for instance, $d\approx4$
at the laboratory or classical scale and may run to $d<4$ fractal
dimensions at other scales. Therefore, although we derive the Ricci
flow equation (\ref{eq:ricci-flow-t}) from the laboratory scale where
$d\approx D=4$, the equation does not specify a particular $d$,
it is an approximate RG flow for a NLSM with $d\lesssim D=4$.

The $d$-insensitivity of the Ricci flow to the Q-NLSM is obviously
an approximation because we know the low energy phase structure of
the NLSM is $d$ dependent. The Ricci flow is just a Gaussian approximation
to the RG flow of NLSM ignoring higher (than 2nd order) moments fluctuations,
which only describes the phase far away from the low energy phase
transition where higher moments fluctuations are important. The high
energy regions in the phase diagrams of NLSM are common for different
$d\lesssim4$ because $\pi_{d}(S^{4})$ are trivial, and no matter
what the effective dimension $d$ is or how it runs, the Ricci flow
as a Gaussian approximation captures their common high energy behaviors.
Fortunately, the quantity (corresponding to CC) we need to calculate
by the method is a high energy limit of Perelman's reduced volume,
and we only concern its convergence at UV, so this shortcoming of
the Ricci flow does not really bother us. When the NLSM is near a
low energy phase transition point, the moments or fluctuation higher
than 2nd order can not be ignored, the quantum fluctuation of coordinates
are non-Gaussian, then terms being composed of higher power $(R_{\mu\nu}/\lambda)^{n}$
also come into the flow. The Ricci flow equation as a Gaussian approximation
fails, such type of flow is beyond the scope of the paper, some literature
show that such type of equations qualitatively behaves similar with
the Ricci flow \cite{guenther2008stability,gimre2012second}. For
instance, for some rescaled Ricci flow equation, the diffusion part
determines the qualitative behavior and the non-Gaussian fluctuations
are irrelevant, then as an initial geometry with non-Gaussian fluctuations
flows, the fluctuation becomes more and more Gaussian, the Ricci flow
describes its correct qualitative behavior.

\section{Spacetime with Intrinsic 2nd Moment as a Quantum Reference System}

The next question naturally is how a to-be-studied quantum system
is described be relative to the spacetime with intrinsic 2nd moment
as a reference system. To interpret the above defined spacetime theory,
it needs to be re-formulated in an effective and semi-classical form
as our familiar spacetime theory in physics (as General Relativity
does).

\subsection{Quantum Entanglement between a to-be-studied system and a reference
system: relational interpretation }

Ordinary textbook quantum mechanics and quantum field theories are
formulated with respect to a classical and absolute parameter background
free from any quantum fluctuations, which is not physical and arises
severely problem when gravity is taken into account, for example the
cosmological constant problem. In a more physical treatment that a
to-be-studied system and a spacetime reference system are both quantum,
the to-be-studied system are described by a state $|\Psi\rangle$
in Hilbert space $\mathscr{H}_{\Psi}$, the Quantum Spacetime Reference
Frame (QSRF) system are described by a state $|X\rangle$ in Hilbert
space $\mathscr{H}_{X}$, and the states of both systems are given
by an entangled state 
\begin{equation}
|\Psi[X]\rangle=\sum_{ij}\alpha_{ij}|\Psi\rangle_{i}\otimes|X\rangle_{j}\label{eq:entangled state}
\end{equation}
in their direct product Hilbert space $\mathscr{H}_{\Psi}\otimes\mathscr{H}_{X}$.
Here the state denoted by $|\Psi[X]\rangle$ is with respect to the
quantum spacetime coordinate $X$, in analogy with a state $|\Psi(x)\rangle$
of the quantum system being with respect to a classical spacetime
coordinate $x$. 

The reason why the state of both systems are in an entangled state
but a direct product state is as follows. Before a quantum measurement
is performed, an important step, called calibration, is always explicitly
or implicitly carried out. At the step, a one-to-one correlation between
a state $|\Psi\rangle_{i}$ of the to-be-studied system and a state
$|X\rangle_{j}$ of the quantum measuring instrument must be established.
Calibration in usual sense is a comparison and adjustment of the measuring
instrument $|X\rangle_{j}$ by a calibration standard $|\Psi\rangle_{i}$
of a standard system which is physically the same with the to-be-studied
system, namely when the standard system is in the state $|\Psi\rangle_{i}$
the measuring instrument must be adjusted to be in state $|X\rangle_{j}$.
This process of adjustment introduces an entangled state $\sum_{ij}\alpha_{ij}|\Psi\rangle_{i}\otimes|X\rangle_{j}$
which describes the state $|\Psi\rangle_{i}$ with respect to a quantum
measuring instrument $|X\rangle_{j}$. 

The entangled state $\sum_{ij}\alpha_{ij}|\Psi\rangle_{i}\otimes|X\rangle_{j}$
is a generalization of a textbook quantum state $|\Psi\rangle_{i}$
which is with respect to a classical measuring instrument. In this
sense, $\Psi[X]$ being a functional is a generalization of $\Psi(x)$
being a function. The functional $\Psi[X]$ can be seen as a function
$\Psi$ with a smeared variable $X$. And after that calibration step,
the calibration standard is removed away, and every time we read the
state $|X\rangle_{j}$ of the measuring instrument, in the usual sense,
it infers to a state $|\Psi\rangle_{i}$ of the to-be-studied system
according to the established entangled correlation. Now the measuring
instrument, for instance rulers and clocks measuring the space and
time coordinates $X$ as a reference \cite{PhysRevD.27.2885,2014PhRvA..89e2122M},
and the to-be-studied system are both quantum. 

Taking the Stern-Gerlach experiment as a simple example, which measures
the angular momentum or spin of a particle. A particle (the to-be-studied
system) taking spin state $|\Psi\rangle_{1}=|\uparrow\rangle$ or
$|\Psi\rangle_{2}=|\downarrow\rangle$ passes through a non-uniform
magnetic field, and hits a detector screen. The state of the screen
(the measuring instrument) is either $|X\rangle_{1}=|+\rangle$ if
the upper-half-plane of the screen is hit and lights up or $|X\rangle_{2}=|-\rangle$
if the lower-half-plane of the screen is hit and lights up. Before
the experiment is performed, we must according to the known law of
how a spin moves in a non-uniform magnetic field to adjust the screen
or the magnetic field, making that the particle in state $|\uparrow\rangle$
triggers the upper-half-plane of the screen and $|\downarrow\rangle$
state triggers the lower-half-plane of the screen. In other words,
in the step of experiment calibration, we prepare an entangled state
$\frac{1}{\sqrt{2}}|\uparrow\rangle|+\rangle+\frac{1}{\sqrt{2}}|\downarrow\rangle|-\rangle$
to the whole system and establish a one-to-one correlation between
the state of the particle spin and the detector screen. When the experiment
is carried out, a particle of unknown spin passes through the non-uniform
magnetic field, if we see the state of the screen is $|+\rangle$
then we know the spin of the particle is $|\uparrow\rangle$, and
similarly if the state of the screen is $|-\rangle$ the spin of the
particle is $|\downarrow\rangle$.

In the standard interpretation of quantum state, a measuring of $|X\rangle$
can spookily collapse $|\Psi\rangle$, even though one has not touch
it at distance. So measuring $|X\rangle$ tells you some information
about $|\Psi\rangle$. More precisely, the expanding coefficient $\alpha_{ij}$
gives the amplitude of the to-be-studied system being in state $|\Psi\rangle_{i}$
and the measuring instrument being in state $|X\rangle_{j}$. $|\alpha_{ij}|^{2}$
is the joint probability related to the amplitude. The amplitude $\alpha_{ij}$
of the entangled state in general can not be factorized into a product
of each individual absolute amplitude of $|\Psi\rangle_{i}$ and $|X\rangle_{j}$,
which called ``non-separability of entangled state''. In this sense,
The essential thing is not the individual states $|\Psi\rangle_{i}$
or $|X\rangle_{j}$ but their relation described by their entanglement,
in other words, the interpretation of the state must be relational
\cite{1996IJTP...35.1637R}. In the example of Stern-Gerlach experiment,
we see that the statement ``particle is spin up/down'' has no physical
meaning unless one really clears what ``up/down'' be relative to.
In fact, we see that ``up/down'' is relative to the upper or lower
half-plane of the detector screen. Without the screen, the notion
``up/down'' has no absolute meaning. The relational nature of entangled
state is very important. The state of the system $|\Psi\rangle_{i}$
makes sense only with respect to the state of the instrument $|X\rangle_{j}$
, that is the spirit of relativity and what a reference system $|X\rangle_{j}$
is used for. Only when the reference system becomes classical, in
other words, the 2nd or higher moment of the spacetime coordinates
vanish, the coordinates of the spacetime can be seem as a Dirac's
delta distribution for the state $|X\rangle_{j}$ without any quantum
fluctuation, the relational amplitude $\alpha_{ij}$ reproduces the
standard absolute amplitude $|\Psi\rangle_{i}$ in textbook quantum
mechanics. Taking also the Stern-Gerlach experiment as an example,
this situation corresponding to the states of the screen $|+\rangle$
and $|-\rangle$ have no overlap and hence the spin state can be classically
distinguishable. In general, the reference system at least has non-trivial
2nd moment, it is equivalent to a wavefunction $\Psi$ with blurry
coordinates $X$ \cite{phillips2000vacuum}.

\subsection{Action of the to-be-studied system and the reference system (spacetime)}

After the step of calibration between the to-be-studied system and
the reference system is established and before their next interaction
(i.e. measurement), the two systems may evolve independently without
any interaction. The action of the entangled systems is a direct sum
of each individual actions without interaction. Without loss of generality,
we consider a conventional scalar field $\Psi(x)$ as the to-be-studied
system. The reference system is the spacetime coordinate system given
by rulers $X^{i}(x)$ and clocks $X^{0}(x)$ from the action of NLSM,
which shares the base space $x$ as a common background with the scalar
field $\Psi(x)$. The action is
\begin{equation}
S[\Psi,X^{\mu}]=\int d^{d}x\left[\left(\frac{1}{2}\frac{\partial\Psi}{\partial x^{a}}\frac{\partial\Psi}{\partial x^{a}}-V_{p}(\Psi)\right)+\frac{1}{2}\lambda g_{\mu\nu}\frac{\partial X^{\mu}}{\partial x^{a}}\frac{\partial X^{\nu}}{\partial x^{a}}\right],\label{eq:entangled-action}
\end{equation}
where $V_{p}(\Psi)$ is a potential term of the scalar field. The
action is formulated by the to-be-studied scalar field $\Psi(x)$
and spacetime fields $X^{\mu}(x)$ with respect to the parameter background
$x^{a}$. The standard quantum mechanics and quantum fields theory
are formulated requiring certain inertial frame or parameter background,
for example, quantum mechanics has only one background parameter:
Newton's absolute time, quantum fields theory has four background
parameters: Minkovski's spacetime background $x^{a}$. However, the
parameter background is not necessarily interpreted as the physical
spacetime, because they are absolute, external, classical and free
from any quantum fluctuation. Now the scalar field $\Psi$ must be
described with respect to the physical spacetime $X^{\mu}$ instead
of parameter background $x^{a}$. Since the action describes the entanglement
between the field $\Psi(x)$ and spacetime $X^{\mu}(x)$, the important
thing is not their individual meanings, but their relation, so the
action concerns the relational or topology of the quantum states,
and hence implies a topological quantum field theory relating to certain
``gravity'' theory with a proper interpretation.

\subsection{Mean field approximation: reproduce the classical action}

By the semi-classical or Mean Field (M.F.) approximation in which
only the mean field value $\langle X\rangle$ is considered and its
2nd moment $\langle\delta X^{2}\rangle$ is ignored, it is simply
given by a variable change $x^{a}\rightarrow X^{\mu}=\langle e_{a}^{\mu}\rangle x^{a}$,
the action (\ref{eq:entangled-action}) becomes
\begin{equation}
S[\Psi,X^{\mu}]\overset{M.F.}{=}\int d^{D}X\sqrt{\left|\det g\right|}\left[\frac{1}{4}\left\langle g_{\mu\nu}\frac{\partial X^{\mu}}{\partial x^{a}}\frac{\partial X^{\nu}}{\partial x^{a}}\right\rangle \left(\frac{1}{2}g^{\mu\nu}\frac{\delta\Psi}{\delta X^{\mu}}\frac{\delta\Psi}{\delta X^{\nu}}+2\lambda\right)-V_{p}(\Psi)\right].
\end{equation}
$\sqrt{\left|\det g\right|}$ is the Jacobian determinant of the variable
change. At the mean field approximation, we see that it must have
$d=D\equiv4$, because the Jacobian determinant $\sqrt{\left|\det g\right|}=\left\Vert \frac{\partial x_{a}}{\partial X^{\mu}}\right\Vert $
must be a square matrix. As is discussed before, strictly speaking,
$d$ can be very close to but not precisely to 4. So it is worth stressing
that the conclusion $d=4$ is only valid at the mean field level.
Then we have 
\begin{equation}
S\left[\Psi[X^{\mu}]\right]\overset{M.F.}{=}\int d^{4}X\sqrt{\left|\det g\right|}\left[\frac{1}{2}g^{\mu\nu}\frac{\delta\Psi}{\delta X^{\mu}}\frac{\delta\Psi}{\delta X^{\nu}}-V_{p}(\Psi)+2\lambda\right],\label{eq:m.f.action}
\end{equation}
in which $\frac{1}{4}\left\langle g_{\mu\nu}\frac{\partial X^{\mu}}{\partial x^{a}}\frac{\partial X^{\nu}}{\partial x^{a}}\right\rangle =\frac{1}{4}\left\langle g_{\mu\nu}g^{\mu\nu}\right\rangle =\frac{D}{4}=1$
has been used. So when the quantum fluctuation or 2nd moment of spacetime
reference system is ignored, the action reproduces a textbook scalar
field action in a fixed background of classical Riemannian curved
spacetime up to a constant $2\lambda$, only the original derivative
$\partial\Psi(x)/\partial x^{a}$ is replaced by a functional derivative
$\delta\Psi[X]/\delta X^{\mu}$, $\Psi$ now is interpreted as a functional
of spacetime coordinates $X$ instead of a function of parameter background
$x$. In the language of quantum mechanics, the textbook quantum state
$|\Psi[X^{\mu}]\rangle$ of the system (\ref{eq:m.f.action}) is a
mean field approximation to the entangled state (\ref{eq:entangled state})
of the entangled system (\ref{eq:entangled-action}). We will see
later that when the fluctuation or flow of spacetime is taken into
account, dynamics of gravity and CC emerge in (\ref{eq:m.f.action}).

\section{Ricci Flow from An Isotropic Positive Curvature Initial Spacetime}

Here we consider the action beyond the mean field approximation, that
is to consider the geometry of $g_{\mu\nu}$ flows governed by the
Ricci flow induced by the 2nd moment of the spacetime. For an isotropic
initial condition when the diffusion part of the equation wins, it
tends to smooth out local fluctuations of the geometry. Although the
problem of Ricci flow for general 4-manifolds remains open, for an
isotropic and positive curvature initial condition, the situation
is better understood. In such case, when the Ricci flow equation is
properly normalized (actually normalized by a CC), the flow solution
of 4-manifolds exists for all $t\in(-\infty,0)$ and be singularity
free (singularity does not locate at finite $t$), the curvature gradually
becomes positive, isotropic and uniform, and converges to a manifolds
with finite constant volume ratio. 

\subsection{The density $u$ function and renormalization $Z$,$\tilde{Z}$ functions}

To handle the difficult Ricci flow equation, there is a trick in Ricci
flow literature that introduces an auxiliary function 
\begin{equation}
u(X,t)=\frac{1}{(4\pi t)^{D/2}}e^{-f(X,t)}
\end{equation}
canceling the flow of the measure $d^{D}X$ around point $X$, i.e.
\begin{equation}
\frac{\partial}{\partial t}\left[u\sqrt{\left|\det g_{\mu\nu}(X,t)\right|}\right]=0,\quad\mathrm{or\quad}\frac{\partial}{\partial t}\left[ud^{D}X(t)\right]=0.\label{eq:constraint}
\end{equation}
In mathematics the $u(X,t)$ function is often called the density
of a manifolds.

The density function $u(X,t)$ is useful because, firstly, it provides
a fixed measure along the flow, secondly, it just modifies the Ricci
flow equation by a family of diffeomorphism equivalent to the standard
Ricci flow via the scalar function $f(X,t)$, and thirdly, the modified
Ricci flow equation (called Ricci-DeTurck flow \cite{deturck1983deforming})
\begin{equation}
\frac{\partial g_{\mu\nu}}{\partial t}=-2R_{\mu\nu}-2\nabla_{\mu}\nabla_{\nu}f
\end{equation}
turns out to be a gradient flow of some functionals. The variational
structure of Ricci flow was discovered by Perelman.

Here we discuss its meaning in the language of NLSM, if we consider
the flow of an isotropic metric is as 
\begin{equation}
g_{\mu\nu}(X,t)=Z(X,t)g_{\mu\nu}(X,t_{0})\label{eq:Z}
\end{equation}
and hence the local volume flows as 
\begin{equation}
\sqrt{\left|\det g_{\mu\nu}(X,t)\right|}=Z^{D/2}(X,t)\sqrt{\left|\det g_{\mu\nu}(X,t_{0})\right|},
\end{equation}
in which $t_{0}$ is certain starting scale that ordinary physical
and volume measurements are relative to, e.g. $t_{0}$ be the scale
of a Laboratory Wall Frame (LWF) that can be used very precisely as
a reference to align the coordinate frame. Since $u$ cancels the
volume flow, we have the relation
\begin{equation}
u(X,t)=Z^{-D/2}(X,t).
\end{equation}
Thus in fact the invariant measure $ud^{D}X$ is nothing but the measure
of the base space of the NLSM, 
\begin{equation}
ud^{D}X(t)=u\sqrt{\left|\det g_{\mu\nu}(X,t)\right|}d^{D}X(t_{0})=uZ^{D/2}\sqrt{\left|\det g_{\mu\nu}(X,t_{0})\right|}d^{D}X(t_{0})=\sqrt{\left|\det g_{\mu\nu}(X,t_{0})\right|}d^{D}X(t_{0})=d^{4}x,
\end{equation}
i.e. the volume element of LWF as a standard measure is considered
not flow. The function $u(X,t)$ can be seen rescale the geometry
$g_{\mu\nu}(X,t)$ isotropically at each point $X$ at scale $t$
and keeping the volume of the base space fixed. Now since $Z(X,t)=u(X,t)^{-2/D}$
can be interpreted as an isotropic flow of the NLSM action
\begin{equation}
S_{X}(t)=\frac{1}{2}\lambda\int d^{d}xZg_{\mu\nu}\partial_{a}X^{\mu}\partial_{a}X^{\nu}=\frac{1}{2}\lambda\int d^{d}xu^{-2/D}g_{\mu\nu}\partial_{a}X^{\mu}\partial_{a}X^{\nu}.\label{eq:renormalized NLSM}
\end{equation}
The flow can also be interpreted as a uniform flow of the parameter
$\lambda$, i.e. $\lambda_{t}=Z(t)\lambda$, if $Z(X,t)\approx Z(t)$
is weakly depends on the coordinate $X$ and thus be able to take
out of the integral, while other quantities are considered fixed.

The flow of the volume in Ricci flow is a forwards heat-like equation,
and since the flow of $u$ cancels the volume flow, it is given by
a backwards heat-like equation on the manifolds, from (\ref{eq:constraint})
we have 
\begin{equation}
\boxempty^{*}u=\left(-\frac{\partial}{\partial t}-\Delta+R\right)u=0,\label{eq:conjugate-eq-u}
\end{equation}
where $\Delta$ is the Laplace\textendash Beltrami operator of the
$D$-dimensional spacetime and $\boxempty^{*}$ is conjugate to the
heat operator $\boxempty=\frac{\partial}{\partial t}-\Delta$. 

Since the flow of $u$ completely describes the flow of volume induced
by the isotropic flow of metric, so for the isotropic case, we study
the flow of $u$ on the flowing manifolds instead of directly studying
the difficult multi-component Ricci flow, it simplifies the problem
we deal with. Because different choices of $u$ lead to the same Ricci
flow up to diffeomorphisms, so the choice of it is analogous to the
choice of gauge. We can see obviously that multiplying $u$ by a constant
just gives the action (\ref{eq:renormalized NLSM}) an unimportant
redefinition and leaves the Ricci-DeTurck flow unchanged. This means
that one can choose $u$ by convenience and it is not unique, it only
provides a convenient tool instead of introducing an extra scalar
field to the theory. However, in practice, if we impose a starting
condition to its flow equation, it can be determined uniquely by solving
the flow equation. The partial goal of the following chapter is to
determine its IR value $u_{0}$ by imposing a UV renormalization condition. 

Naively, the backwards heat flow (\ref{eq:conjugate-eq-u}) will not
exist for general $u$. However, one of the basic points of view is
to let the Ricci flow go for a IR $t_{*}\ge0$. At $t_{*}$ one may
then choose an appropriate $u(t_{*})=u_{0}$ arbitrarily and flow
it backwards in $t$ ($\tau=t_{*}-t$) to obtain a solution $u(t)$
of the backwards equation. Since for the isotropic case, the flow
is free from singularity, it simply gives $t_{*}=0$, so we define
\begin{equation}
\tau=0-t=\frac{1}{64\pi^{2}\lambda}k^{2}.\label{eq:tau-k}
\end{equation}
It reverses the direction of $t$ and making the flow of $u$ a more
familiar forwards heat-like equation
\begin{equation}
\frac{\partial u}{\partial\tau}=\Delta u-Ru,\label{eq:eq-u}
\end{equation}
which does admit a solution in $\tau$ in the sense discussed above,
and equivalently, a backwards solution in $t$.

In a fixed metric with a small scalar curvature, it can be considered
become a heat equation. Thus we expect the fundamental solution forms
almost like a standard heat kernel 
\begin{equation}
H(X;Y,g,\tau)=\frac{1}{(4\pi\tau)^{D/2}}e^{-\frac{|X-Y|^{2}}{4\tau}}.\label{eq:heat kernel}
\end{equation}
The information of the curvature can be reflected in the Fourier transformation
of $H(X;Y,g,\tau)$ from $X$ space to its momentum space $K$
\begin{equation}
\tilde{u}(K,\tau)=\int_{M^{D}}H(X;Y,g,\tau)e^{iK_{\mu}(X^{\mu}-Y^{\mu})}d^{D}X=e^{-|K|^{2}\tau},
\end{equation}
and the Fourier transformation of $Z(X,\tau)$ is given by 
\begin{equation}
\tilde{Z}(K,\tau)=e^{\frac{2}{D}|K|^{2}\tau}\approx1+\frac{2}{D}R\tau.\label{eq:fourier-Z}
\end{equation}
It renormalizes the Fourier components of the metric
\begin{equation}
g_{\mu\nu}(K,\tau)=\tilde{Z}(K,\tau)g_{\mu\nu}(K),\label{eq:definition tilde Z}
\end{equation}
where $|K|^{2}$ is the eigenvalue of the operator $-\Delta+R\approx R$
when $\tilde{Z}$ is uniform enough.

The definition domain of $u$ can be even $\tau\in[0,\infty)$, called
ancient solution in mathematics literature, if the existence of the
solution can be traced back to $\tau\rightarrow\infty$, in physics
it means that the system exists a UV fixed point $k\rightarrow\infty$.
It is indeed the case for the isotropic and positive curvature spacetime
because it is singularity free. So we assume 
\begin{equation}
u_{\infty}^{-2/D}=Z_{\infty}=1\label{eq:u_inf}
\end{equation}
to be the final condition of (\ref{eq:eq-u}), which is actually a
renormalization condition at UV. The existence of $u_{\infty}$ is
called renormalizability in physics. $u_{\infty}=1$ means that the
action (\ref{eq:NLSM}) is considered as a bare action with respect
to (\ref{eq:renormalized NLSM}). Our goal is to estimate the initial
$u$ and $\tilde{Z}$ at IR $\tau\rightarrow0$ by the renormalization
condition at UV, then we could finally obtain the IR limit of the
action.

\subsection{Estimate initial density $u_{0}$ by the $v$ function and the reduced
volume}

In order to study the solution $u$ of the equation (\ref{eq:eq-u}),
Perelman introduced a subsolution \cite{perelman2002entropy,topping2006lectures,muller2006differential}
\begin{equation}
v=\left[\tau\left(R+2\Delta f-|\nabla f|^{2}\right)+f-D\right]H,
\end{equation}
to the conjugate heat equation (\ref{eq:conjugate-eq-u}) 
\begin{equation}
\boxempty^{*}v=-2\tau\left|R_{\mu\nu}+\nabla_{\mu}\nabla_{\nu}f-\frac{1}{2\tau}g_{\mu\nu}\right|^{2}H\le0\label{eq:eq-v}
\end{equation}
as a useful tool to estimate the fundamental solution $H(X;Y,g,\tau)$
of (\ref{eq:eq-u}). The inequality holds as equality when the manifolds
is a gradient shrinking soliton solution satisfying
\begin{equation}
R_{\mu\nu}+\nabla_{\mu}\nabla_{\nu}f-\frac{1}{2\tau}g_{\mu\nu}=0.\label{eq:soliton eq}
\end{equation}
The fundamental subsolution $v$ is the lower bound of the fundamental
solution $H$ 
\begin{equation}
H(X;Y,g,\tau)\ge v(X;Y,g,\tau)=\frac{1}{(4\pi\tau)^{D/2}}e^{-l(X-Y,g,\tau)},\label{eq:u>=00003Dv}
\end{equation}
where $l(X;Y,g,\tau)$ is the reduced length \cite{perelman2002entropy}
measuring the minimum ``distance'' of a path between the base point
$\gamma(0)=Y$ and end point $\gamma(\tau)=X$, 
\begin{equation}
l(X;Y,g,\tau)=\inf_{\gamma}\frac{1}{2\sqrt{\tau}}\int_{0}^{\tau}\sqrt{\tau^{\prime}}\left[R\left(\gamma(\tau^{\prime})\right)+\left|\frac{d\gamma}{d\tau^{\prime}}\right|_{g(\tau^{\prime})}^{2}\right]d\tau^{\prime},
\end{equation}
where $R\left(\gamma(\tau^{\prime})\right)$ is the scalar curvature
at the point $\gamma(\tau^{\prime})$ and the $\inf$ (infimum) is
taken over all paths $\gamma$ with fixed base and end points. The
volume integral,
\begin{equation}
\bar{V}(X,g,\tau)\equiv\int_{M^{D}}d^{D}Yv(X;Y,g,\tau)\equiv\int_{M^{D}}d^{D}Y\frac{1}{(4\pi\tau)^{D/2}}e^{-l(X-Y,g,\tau)}\label{eq:RV}
\end{equation}
is called the reduced volume \cite{perelman2002entropy} of the spacetime
$M^{D}$ with basepoint $X$. The reduced volume is a dimensionless
geometric quantity and monotone non-increasing in $\tau$ which even
holds on noncompact manifolds. Perelman used the quantity to prove
the no-local-collapsing theorem of a manifolds via the monotonic of
it.

The reduced volume is a generalization of a volume integral of the
heat-kernel (\ref{eq:heat kernel}), for example, the integral of
(\ref{eq:heat kernel}) in a D-dimensional flat manifolds gives
\begin{equation}
\int_{\mathbb{R}^{D}}d^{D}Y\frac{1}{(4\pi\tau)^{D/2}}e^{-\frac{|X-Y|^{2}}{4\tau}}=1.\label{eq:RV_R^D}
\end{equation}
And the reduced length $l(X;Y,g,\tau)$ is a variant of the Gaussian
distance $f(X;Y,g,\tau)=|X-Y|^{2}/4\tau$ up to a constant. We see
that (\ref{eq:RV_R^D}) leads to a heuristic
\begin{equation}
H(X;Y,g,\tau)=\frac{1}{(4\pi\tau)^{D/2}}e^{-\frac{|X-Y|^{2}}{4\tau}}\approx\frac{1}{\mathrm{Vol}(\mathbb{R}^{D})}.
\end{equation}
Thus the main difference between the reduced volume in $M^{D}$ and
in $\mathbb{R}^{D}$ comes from the difference in the integrals of
volumes, by taking $H\ge v$, the (\ref{eq:RV}) is then 
\begin{equation}
\bar{V}(g,\tau)\le\frac{\mathrm{Vol}(M^{D},g(\tau))}{\mathrm{Vol}(\mathbb{R}^{D})}.\label{eq:volume ratio}
\end{equation}
Therefore, the reduced volume is bounded above by the volume ratio,
which is well-defined also due to the Bishop-Gromov comparison theorem,
and for general $M^{D}$ we always have 
\begin{equation}
\bar{V}(\tau)\le1.
\end{equation}

Here we discuss a useful observation for $H$ and $v$ on an isotropic
and positive curvature manifolds having a backwards UV limit. On the
one hand, since for positive curvature manifolds, its IR limit is
a gradient shrinking soliton satisfying (\ref{eq:soliton eq}) and
hence (\ref{eq:u>=00003Dv}) holds as equality at $\tau\rightarrow0$.
On the other hand, since the backwards existence of the solution is
able to extend to $\tau\rightarrow\infty$, i.e. an ancient solution.
There are subtle interplay between ancient solution and soliton, a
rescaled ancient solution resembles a soliton, more precisely, its
backwards limit is an asymptotic non-flat gradient shrinking soliton
\cite{perelman2002entropy,Cao2010The} also satisfying (\ref{eq:soliton eq}),
in this situation the equality (\ref{eq:eq-v}) also holds at UV.
As a consequence, $H$ equals $v$ up to a constant multiple when
the solution of Ricci flow is an ancient solution. The constant multiple
for the fundamental solution is not important since it can be finally
fixed by imposing an initial condition to the general solution $u$.
In fact the ratio 
\begin{equation}
W(\tau)=\frac{v}{H}=\tau\left(R+2\Delta f-|\nabla f|^{2}\right)+f-D,
\end{equation}
whose IR and UV limits are the constant multiples, defines a local
entropy \cite{perelman2002entropy} which is monotone non-decreasing
in $t=-\tau$. If we start the Ricci flow from an isotropic gradient
shrinking soliton at UV the local entropy $W(\infty)$ has already
taken its maximum constant value equaling to the maximum local entropy
$W(0)$ for the IR gradient shrinking soliton, which can be set to
1 by a normalized $f$ (adding an appropriate constant). So $H$ must
be equal to $v$ at both IR and UV limits for the ancient solution
we concern, and because of their monotonicity, this immediately implies
that $H(X;Y,g,\tau)=v(X;Y,g,\tau)$ for all $\tau\in[0,\infty)$ (instead
of being a subsolution).

As the forward flow makes the density $u$ more and more homogeneous
and isotropic at IR, we set the IR initial condition a constant density
$u_{0}=\mathrm{const}$, and using the fundamental solution $H(X;Y,g,\tau)$
of (\ref{eq:eq-u}) we write its general solution as 
\begin{equation}
u(X,g,\tau)=\int_{M^{D}}d^{D}Yu_{0}H(X;Y,g,\tau).
\end{equation}
In the IR limit the fundamental solution $H$ tends to a delta function,
$\lim_{\tau\rightarrow0}H(X;Y,g,\tau)=\delta^{(D)}(X-Y)$ , so $\lim_{\tau\rightarrow0}u(X,g,\tau)=u_{0}$,
and using the equality between $H$ and $v$ at $\tau\rightarrow0$

\begin{equation}
u_{0}=\lim_{\tau\rightarrow0}\int d^{D}Yu_{0}H(X;Y,g,\tau)=\lim_{\tau\rightarrow0}\int d^{D}Yu_{0}v(X;Y,g,\tau)=u_{0}\bar{V}_{0}(X),
\end{equation}
so we have $\bar{V}_{0}(X)=1$ as expected. And in the UV limit, by
using the renormalization condition (\ref{eq:u_inf}) and the equality
between $H$ and $v$ also at $\tau\rightarrow\infty$, we have

\begin{equation}
1=u_{\infty}(X)=\lim_{\tau\rightarrow\infty}\int d^{D}Yu_{0}H(X;Y,g,\tau)=\lim_{\tau\rightarrow\infty}\int d^{D}Yu_{0}v(X;Y,g,\tau)=u_{0}\bar{V}_{\infty}(X),
\end{equation}
then we obtain 
\begin{equation}
u_{0}=\bar{V}_{\infty}^{-1},\label{eq:u0}
\end{equation}
independent to the basepoint.

This is a basic result of the paper, it shows that the homogeneous
IR initial density $u_{0}$ is determined by the UV limit of the reduced
volume $\bar{V}_{\infty}=\lim_{\tau\rightarrow\infty}\bar{V}(g,\tau)<1$.
It is because the subtle relation between IR and UV that a rescaled
IR limit solution with positive curvature resembles an ancient solution
whose backwards UV limit converges to a non-flat gradient shrinking
soliton. And since the equality in (\ref{eq:eq-v}) holds at UV, the
UV reduced volume given by $v$ fixes the initial density $u_{0}$.

For general $\tau\ge0$ we have
\begin{equation}
u(X,g,\tau)=\int_{M^{D}}d^{D}Yu_{0}H(X;Y,g,\tau)=\int_{M^{D}}d^{D}Yu_{0}v(X;Y,g,\tau)=u_{0}\bar{V}(X,g,\tau),
\end{equation}
so the reduced volume for all $\tau$ is given by 
\begin{equation}
\bar{V}(X,g,\tau)=\frac{u(X,g,\tau)}{u_{0}}=u(X,g,\tau)\bar{V}_{\infty},\label{eq:RV flow}
\end{equation}
 which shows that the reduced volume is in fact monotonic decreasing
in $\tau$ as $u(X,g,\tau)$ behaves.

In the language of NLSM, remind that the standard volume of the base
space is just the invariant volume constraint of the target space,
\begin{equation}
\int_{base}d^{4}x=\int_{target}u(\tau)d^{D}X=\lim_{\tau\rightarrow0}\int_{target}d^{D}X\int_{target}d^{D}Yu_{0}H(X;Y,g,\tau)=\bar{V}_{\infty}^{-1}\int_{target}d^{D}X(0).
\end{equation}
Thus in the sense of comparison geometry, the UV reduced volume is
just a volume ratio between IR volume of the target space and standard
volume of the base space
\begin{equation}
\bar{V}_{\infty}=\frac{\int_{target}d^{D}X(0)}{\int_{base}d^{4}x}<1.\label{eq:uv rv}
\end{equation}
It means that if the flow of spacetime starts from a small distance
scale (e.g. LWF) forwardly to the long distance scale, the Ricci flow
shrinks the spacetime volume and finally converges to a constant volume
at IR limit.

The convergence of $\bar{V}_{\infty}$ and $u_{0}$ are crucial facts
which relate to the renormalizability and asymptotic safety of the
Q-NLSM, leading to a convergent value of CC. Actually, $\bar{V}_{\infty}(g)$
is the Gaussian density $\Theta(g)$ \cite{yokota2010asymptotic,cao2004gaussian,cao2009recent}
of a UV manifolds, its ``$\mathit{\ln}$'' can be given by a limit
of the $\mathcal{W}$-functional introduced also by Perelman \cite{perelman2002entropy,feldman2005entropy,xu2017equation},
\begin{equation}
\ln\bar{V}_{\infty}(g)=\lim_{\tau\rightarrow\infty}\mathcal{W}(g,f,\tau)=\mathcal{W}_{\infty}(g)<0,
\end{equation}
where
\begin{equation}
\mathcal{W}(g,f,\tau)=\int_{M^{D}}\left[\tau\left(R+|\nabla f|^{2}\right)+f-D\right]ud^{D}X
\end{equation}
is monotone non-increasing in $\tau$, in other words, it is monotone
non-decreasing along the Ricci flow like an entropy.

At the point the initial condition is given by $u_{0}=e^{-\mathcal{W}_{\infty}}$,
and imposing the renormalization condition (\ref{eq:u_inf}), $\tilde{Z}(K,\tau)$
becomes a normalized version of (\ref{eq:fourier-Z})
\begin{equation}
\tilde{Z}(K,\tau)=e^{\frac{2}{D}\nu+\frac{2}{D}|K|^{2}\tau}\approx1+\delta_{\tilde{Z}}+\frac{2}{D}R\tau,\label{eq:tilde_Z}
\end{equation}
where $\delta_{\tilde{Z}}=\tilde{Z}(K,0)-1\approx\frac{2}{D}\mathcal{W}_{\infty}$
is a counter term completely canceling $\frac{2}{D}R\tau$ at $\tau\rightarrow\infty$.
By substituting (\ref{eq:tau-k}), it coincides with the result in
ref.\cite{Luo:2015pca} leaving the constant $\delta_{\tilde{Z}}$
to be determined latter. 

Let us summarize this section, the result (\ref{eq:RV flow}) shows
that the density $u$ starting from $u_{0}=\bar{V}_{\infty}^{-1}>1$
at IR flows backwardly to $u_{\infty}=1$ at UV, due to the fact that
the reduced volume starting from $\bar{V}_{0}=1$ flows backwardly
to $\bar{V}_{\infty}<1$, both processes are monotone non-increasing
in $\tau$. As a consequence, the function $\tilde{Z}$ starting from
$\tilde{Z}_{\infty}=1$ at UV flows forwardly to an IR value $\tilde{Z}_{0}$
related to $\bar{V}_{\infty}$ as follows 
\begin{equation}
\tilde{Z}_{0}=Z_{0}=u_{0}^{-\frac{2}{D}}=\bar{V}_{\infty}^{\frac{2}{D}}=\Theta^{\frac{2}{D}}=e^{\frac{2}{D}\mathcal{W}_{\infty}}<1.\label{eq:Z-u-V}
\end{equation}
If $\tilde{Z}$ is interpreted as a renormalization of the UV bare
parameter $\lambda$ while other quantities such as metric being fixed,
it uniformly flows from $\lambda$ at UV to $\tilde{Z}_{0}\lambda<\lambda$
at IR.

\subsection{Asymptotic UV reduced volume of a maximally symmetric spacetime (late
epoch universe)}

We have seen that $\delta_{\tilde{Z}}$ is just a counter term for
a global and isotropic Ricci flow $\frac{2}{D}R\tau$ in (\ref{eq:tilde_Z}).
To calculate $\delta_{\tilde{Z}}$ and related $\bar{V}_{\infty}$,
we need to flow the reduced volume $\bar{V}(g,\tau)$ from a initial
spacetime to the UV limit. A physical choice of initial condition
for such flow at current or late epoch is a maximally symmetric spacetime,
for instance, a Friedman-Robertson-Walker (FRW) spacetime $M^{D}=\mathbb{R}\times S^{3}$,
which is an isotropic and homogeneous spacetime geometry with a positive
curvature: 
\begin{equation}
ds^{2}=(d\mathcal{T})^{2}-a^{2}(\mathcal{T},\tau)(d\Sigma_{3})^{2},\label{eq:FRW}
\end{equation}
with $a(\mathcal{T},\tau)$ the scale factor or radius of $S^{3}$.
The metric takes an ansartz of a noncompact gradient shrinking soliton
which resembles a 4-dimensional round cylinder. Since here we concentrate
on the CC corresponding to the late epoch universe $\mathcal{T}=T_{0}$,
when the space and time are on an equal footing and symmetric in scales,
the metric can be formally rewritten as an isotropic 4-ball of radius
$a(T_{0},\tau)$ with spacetime origin $\{0\}$ removed: $(0,a(T_{0},\tau)]\times S^{3}\approx\mathbb{B}^{4}-\{0\}$.
Since $T_{0}$ is a fixed late physical time, the radius is isotropic
and homogeneous being not relevant to any spacetime coordinates, so
it can be denoted as $a(T_{0},\tau)\equiv a(\tau)$ for short. Then
the metric looked approximately flat when the isotropic radius is
large at the Late Epoch (L.E.), 
\begin{equation}
ds_{L.E.}^{2}=a^{2}(\tau)\left[(dT)^{2}-(d\Sigma_{3})^{2}\right],\label{eq:late time}
\end{equation}
in which $(d\mathcal{T}){}^{2}=a^{2}(\tau)(dT){}^{2}$ is the rescaled
metric of the temporal distance in the range $\mathbb{R}\in(0,a(\tau)]$,
and $a^{2}(\tau)(d\Sigma_{3})^{2}$ is the rotational symmetric round
metric of the 3-sphere $S^{3}$ with radius $a(\tau)$. Thus at this
point, we want to consider the Ricci flow of an isotropic and homogeneous
universe from the spacetime origin $\{0\}$ extending to the isotropic
maximum radius $a(\tau)$ at the late epoch $T_{0}$, with an approximately
flat and homogeneous scalar function $f(X,\tau)=f(\tau)$. The initial
spacetime of the Ricci flow is obviously different from an anisotropic
initial spacetime of $a(\mathcal{T}\rightarrow0,\tau)$ with a different
scalar function $f(\mathcal{T}\rightarrow0,\tau)$ in the very early
universe when the spatial part may develop a singularity while the
temporal part may be flat, which is not relevant to the late epoch
acceleration and CC, and we will leave the anisotropic case for our
future consideration.

The maximally symmetric initial spacetime (\ref{eq:late time}) with
a specified $a(\tau)$ (to be determined later) is just an Einstein
manifolds. Einstein manifolds as a flow limit is a special soliton
solution, in the sense that the Ricci flow only uniformly dilating
$a(\tau)$ instead of locally deforming the shape of the geometry,
so the Ricci flow keeps the structure of that metric along $\tau$,
only $a(\tau)$ changes its value uniformly with the flow. As a consequence,
the backwards UV limit of the metric does exist and must also be an
Einstein manifolds (\ref{eq:late time}) with merely a different radius
$a(\tau\rightarrow\infty)$, thus (\ref{eq:late time}) as an ancient
solution and a gradient shrinking soliton looks like a 4-ball with
isotropic maximum radius $a(\infty)$ at UV. 

If the chosen initial metric is slightly not such maximally symmetric
form (\ref{eq:late time}) due to local inhomogeneity and anisotropy,
it is naturally considered when the local inhomogeneity and anisotropy
are not large enough so that they will gradually be smoothed out by
the Ricci flow and finally the metric tends to that form, so we can
always find an initial metric sufficiently close to (\ref{eq:late time})
so that its backwards flow limit at UV $\tau\rightarrow\infty$ must
also be that form because of the merely rescaling nature of the soliton's
flow. The goal here is to calculate the UV limit of the reduced volume
by the soliton metric and hence its Gaussian density. 

The radius $a(\tau)$ of the isotropic 4-ball satisfies the shrinking
soliton equation, i.e. a Einstein manifolds of positive curvature
(diffeomorphic normalized by $\nabla_{\mu}\nabla_{\nu}f(\tau)=0$)
\begin{equation}
R_{\mu\nu}=\frac{D-1}{a^{2}}g_{\mu\nu}=\frac{1}{2\tau}g_{\mu\nu},\label{eq:shrinking}
\end{equation}
so we have a shrinking radius $a^{2}(\tau)=2(D-1)\tau$ as $\tau\rightarrow0$.
By taking the maximum radius $a(\tau)$ with $\tau=\frac{a^{2}}{2(D-1)}\rightarrow\infty$,
and according to (\ref{eq:RV}), we obtain
\begin{eqnarray}
\bar{V}_{\infty}\left((0,a(\tau)]\times S^{D-1}\right) & \approx & \int_{\mathbb{B}_{\infty}^{D}-\{0\}}d^{D}X\frac{1}{(4\pi\tau)^{D/2}}e^{-l_{\infty}}\nonumber \\
 & \approx & \left[\frac{D-1}{2\pi a^{2}(\infty)}\right]^{D/2}\int_{0_{+}}^{a(\infty)}e^{-\frac{r^{2}}{4\tau}}(D\omega_{D})r^{D-1}dr\nonumber \\
 & \overset{D=4}{=} & 0.442,
\end{eqnarray}
in which $\mathbb{B}_{\infty}^{D}$ is a $D$-ball in the metric (\ref{eq:late time})
at $\tau\rightarrow\infty$ with a large $a(\infty)\gg1$, $D\omega_{D}=D\pi^{\frac{D}{2}}/\Gamma(\frac{D}{2}+1)$
is the surface area of the unit $D$-ball or the volume of the unit
$(D-1)$-sphere, and $l_{\infty}\approx\frac{r^{2}}{4\tau}$ is approximately
a Gaussian reduced length because of its large scale factor and small
curvature. According to (\ref{eq:uv rv}) the reduced volume $\bar{V}_{\infty}$
measures the asymptotic IR relative volume, i.e. the volume ratio
between $\mathrm{Vol}\left((0,a(\tau\rightarrow0)]\times S_{\tau\rightarrow0}^{3}\approx\mathbb{B}_{\tau\rightarrow0}^{4}\right)$
and standard $\mathrm{Vol}\left(\mathbb{R}^{4}\right)$, which means
that the volume of the spacetime at large distance scale is asymptotically
$\bar{V}_{\infty}\approx0.442$ times of the standard volume of the
observer's LWF. By using (\ref{eq:Z-u-V}) the volume ratio can be
transformed to the isotropic metric ratio, the final results are
\begin{equation}
\tilde{Z}_{0}=\bar{V}_{\infty}^{2/D}=0.663,\quad\delta_{\tilde{Z}}=\tilde{Z}_{0}-1=-0.337.\label{eq:central}
\end{equation}

These are central numerical results of the paper. The counter term
$\delta_{\tilde{Z}}$ (relating to CC will be discussed latter) is
calculated by the UV reduced volume from a maximally symmetric noncompact
spacetime. The IR limit quantity $\tilde{Z}_{0}$ essentially renormalizes
the Ricci flow of the spacetime metric so that it converges at IR
$\tau\rightarrow0$ globally to a spacetime with finite relative volume
(volume ratio) instead of shrinking to a singular point. The value
of $\delta_{\tilde{Z}}$ arising approximately as the counter term
cancels the critical exponent making the density $u$ converge to
the renormalization condition (\ref{eq:u_inf}).

\subsection{Effective action}

To get a precise physical interpretation to the above results, especially
the relation between $\delta_{\tilde{Z}}$ and CC, here it is convenient
to think of the metric as being fixed, and the flow of $\tilde{Z}$
is interpreted as a uniform flow affecting only $\lambda$ in (\ref{eq:renormalized NLSM})
\begin{equation}
\lambda_{\tau}=\tilde{Z}(\tau)\lambda=\lambda\left(1+\delta_{\tilde{Z}}+\frac{2}{D}R_{\tau}\tau\right),
\end{equation}
Replacing $\lambda$ by $\lambda_{\tau}$ in the effective action
(\ref{eq:m.f.action}) and using (\ref{eq:tau-k}) we have
\begin{equation}
S_{k}=\int d^{4}X\sqrt{\left|\det g\right|}\left[\mathcal{L}_{M}+2(1+\delta_{\tilde{Z}})\lambda+\frac{R_{k}}{16\pi^{2}D}k^{2}\right],\label{eq:effective action}
\end{equation}
in which we have denoted $\mathcal{L}_{M}=\frac{1}{2}g^{\mu\nu}\frac{\delta\Psi}{\delta X^{\mu}}\frac{\delta\Psi}{\delta X^{\nu}}-V_{p}(\Psi)$
for the Lagrangian of the matter field in (\ref{eq:m.f.action}).
In the action, $2(1+\delta_{\tilde{Z}})\lambda$ can be interpreted
as an IR constant vacuum energy density and $\lambda$ is a unique
input bare constant of the NLSM and the theory. $\frac{R_{k}}{16\pi^{2}D}k^{2}$
is a flow term coming from the flow of $\tilde{Z}$ reflecting the
quantum correction at scale $k$ on the background metric, the dynamic
of spacetime or gravity comes from this term. 

We consider the action (\ref{eq:effective action}) as an effective
action of matter $\Psi$ field coupling with gravity, so it must recover
the action of matter field coupling with the standard Einstein-Hilbert
(EH) action. A major difference between this effective action and
the standard EH action is that the action has gradient flow but the
standard EH action does not. The standard EH action without CC is
an action at certain fixed scale much shorter than the cosmic scale
where it is successfully tested, for instance the scale from LWF to
the scale of the solar system. However, at the cosmic scale, the standard
EH action deviates from observations, where CC becomes important.
The correspondence between the effective action (\ref{eq:effective action})
and the standard EH action without CC is as follows. When the cutoff
scale $k$ is large, i.e. at short distance scale even at UV, (\ref{eq:effective action})
recovers the standard EH at the scale, where we know that the flow
term $\frac{R_{k}}{16\pi^{2}D}k^{2}$ almost cancels $2\delta_{\tilde{Z}}\lambda$
in order to satisfy the UV renormalization condition $\tilde{Z}=1$
leaving only $2\lambda$ term; while when the cutoff scale $k$ is
small, i.e. at cosmic scale, the flow term can be neglected, and leaving
$2\delta_{\tilde{Z}}$ together with the short distance $2\lambda$
term. In this sense, the $2\lambda$ term plays the role of an EH
term without CC at short distance scale, and $2\delta_{\tilde{Z}}\lambda$
plays the role of a CC which is important at cosmic scale. 

In the interpretation of $2\lambda$, it can be reformulated by a
constant scalar curvature $R_{0}$ and a constant UV energy scale
$k_{UV}$, most naturally the Planck energy scale $k_{UV}\sim G^{-1/2}$.
So like the standard EH term without CC, for instance, we have $2\lambda=\frac{R_{0}}{16\pi G}$.
To interpret the scalar curvature $R_{0}$ introduced, it is indicated
from observations that at the short distance scale the scalar curvature
$R_{0}$ is small and qualitatively given by the Hubble's constant
at current epoch $H_{0}$, i.e. $R_{0}=D(D-1)H_{0}^{2}=12H_{0}^{2}$.
As a consequence, $\lambda$ is nothing but the Critical Density in
cosmology
\begin{equation}
\lambda=\frac{R_{0}}{32\pi G}=\frac{3H_{0}^{2}}{8\pi G}=\rho_{c}.
\end{equation}
Then the CC term in the action is given by, 
\begin{equation}
2\delta_{\tilde{Z}}\lambda=-0.67\rho_{c}=-\rho_{\Lambda}=\frac{-2\Lambda}{16\pi G}\label{eq:cc}
\end{equation}
with the fraction $\Omega_{\Lambda}=\rho_{\Lambda}/\rho_{c}=-2\delta_{\tilde{Z}}\approx0.67$
consistent with the observations. The fraction $\Omega_{\Lambda}\approx0.67$
estimated by the Ricci flow approximation is close to the result $\Omega_{\Lambda}=2/\pi\approx0.64$
by using the effective dimensional reduction method \cite{Luo:2015pca}.

In this interpretation, $2\lambda+\frac{R_{k}}{16\pi^{2}D}k^{2}$
in the action should play the role of an effective EH term without
CC at cutoff scale $k$, i.e.
\begin{equation}
2\lambda+\frac{R_{k}}{16\pi^{2}D}k^{2}=\frac{R_{k}}{16\pi G},\label{eq:EH}
\end{equation}
which is equivalent to a (short-$\tau$) backwards flow of the scalar
curvature,
\begin{equation}
R_{k}=\frac{R_{0}}{1-\frac{1}{D\pi}Gk^{2}},\quad\mathrm{or}\quad R_{\tau}=\frac{R_{0}}{1-\frac{2}{D}R_{0}\tau}.\label{eq:expanding}
\end{equation}
It is clear that $R_{0}$ can also be interpreted as the IR value
of the scalar curvature, being a homogeneous and isotropic positive
lowest curvature background of the spacetime at cosmic scale. $R_{\tau}$
satisfies 
\begin{equation}
\frac{\partial R_{\tau}}{\partial\tau}=\frac{2}{D}R_{\tau}^{2},\label{eq:R_flow}
\end{equation}
which is a homogeneous backwards diffusion-reaction flow equation
of the scalar curvature when it starts from an IR initial uniform
geometry so that the diffusion term $\Delta R$ is small. The backwards
reaction term means that as the flow goes backwardly to the short
distance scale the quantum fluctuations gradually concentrate its
curvature, consequently, as the flow goes forwardly to the long distance
scale the curvature fluctuations are gradually removed and the spacetime
becomes uniform. Naively speaking, if one calculates the gradient
flow of the standard EH action, it gives a backwards flow to the scalar
curvature whose solution typically may not admit. That is the reason
why the standard EH action in general does not have a gradient flow,
which is known as a weakness of the standard EH action and hence it
is considered not a good starting point of a quantum gravity theory.
However, in the situation we concern the backwards flow solution of
$u$ in (\ref{eq:eq-u}) which does exist shown previously, so a homogeneous
backwards flow solution of the scalar curvature and the effective
EH action induced by the backwards flow of $u$ makes sense. The effective
EH action does have a homogeneous gradient flow unlike the standard
EH action which does not in general. Moreover, the flow of the effective
EH action converges backwardly to UV because of the UV convergence
of $u$. 

At first glance, it seems a contradiction that the flow (\ref{eq:expanding})
or (\ref{eq:R_flow}) shows an expanding behavior conflicting with
the shrinking soliton (\ref{eq:shrinking}). Actually, the effective
curvature $R_{k}$ is indeed shrinking because of the existence of
the counter term $2\delta_{\tilde{Z}}\lambda$. Notice that (\ref{eq:expanding})
without the counter term only shows a short-$\tau$ behavior, and
the counter term is negative $2\delta_{\tilde{Z}}\lambda<0$, so the
effective curvature $R_{k}$ is always positive in its flow. The counter
term calculated from the UV limit of the shrinking soliton (\ref{eq:shrinking})
is able to completely cancel the expanding behavior in (\ref{eq:expanding})
at UV limit. Thus in this sense the counter term $2\delta_{\tilde{Z}}\lambda$
normalizes the flow so that the flow shrinks and converges to a spacetime
with finite volume ratio at IR limit and hence removes the singularity
in (\ref{eq:expanding}).

In the sense that a homogeneous backwards flow solution of the effective
EH action admits and converges at UV, by using (\ref{eq:cc}) and
(\ref{eq:EH}) the effective action (\ref{eq:effective action}) of
gravity can be rewritten as our familiar form
\begin{equation}
S_{k}=\int d^{4}X\sqrt{\left|\det g\right|}\left[\mathcal{L}_{M}+\frac{R_{k}}{16\pi G}+2\delta_{\tilde{Z}}\lambda\right]=\int d^{4}X\sqrt{\left|\det g\right|}\left[\mathcal{L}_{M}+\frac{R_{k}}{16\pi G}-0.67\lambda\right].
\end{equation}

In fact, if one does not introduce the cut off energy scale, the Newton's
constant $G$, the theory with the only input constant $\lambda$
can also be formulated and well-defined in general
\begin{equation}
S_{k}=\int d^{4}X\sqrt{\left|\det g\right|}\left[\mathcal{L}_{M}+2\lambda(\mathfrak{R}_{k}-0.34)\right],\label{eq:effective-action}
\end{equation}
where $\mathfrak{R}_{k}$ is just a dimensionless scalar curvature
equivalent to the conventional scalar curvature $R_{k}$ rescaled
by the IR scalar curvature, i.e. $\mathfrak{R}_{k}=\frac{R_{k}}{R_{0}}=\frac{R_{k}}{32\pi G\lambda}$.
In the context of traditional Einstein's gravity theory with Newton's
constant, the scalar curvature of the theory is seem bound from below
by $R_{0}$ in the isotropic case. By using $\mathfrak{R}$ the resulting
classical field equation is
\begin{equation}
(\mathfrak{R}_{\mu\nu})_{k}-\frac{1}{2}(g_{\mu\nu})_{k}\mathfrak{R}_{k}+0.67(g_{\mu\nu})_{k}=\frac{(T_{\mu\nu})_{k}}{4\lambda},
\end{equation}
which is a rescaled Einstein's equation but Newton's constant plays
no role in it, if we note that $(\mathfrak{R}_{\mu\nu})_{k}\lambda=\frac{(R_{\mu\nu})_{k}}{32\pi G}$.
This equation is in analogy with the Friedman equation $H^{2}/H_{0}^{2}=\rho/\rho_{c}$
in which $\rho$ are densities of matter components rescaled by the
critical density $\rho_{c}$ and the Hubble's parameter $H$ is rescaled
by its current value $H_{0}$ while the Newton's constant is absent
as well.

It is worth noticing that in the theory based on $\mathfrak{R}$,
the critical density $\lambda=\rho_{c}$ is the only characteristic
energy scale instead of the Planck scale. The traditional gravity
theory with a non-vanishing CC has two fundamental constants, the
Newton's constant and the CC, and hence has two characteristic scales,
the Planck scale and Hubble scale. As is shown above, this theory
with only one constant $\lambda$ can also reproduce the traditional
gravity theory with CC by choosing a specific cut off scale, the Planck
scale. However, this theory allows the cut off goes beyond the Planck
scale $k\rightarrow\infty$ while keeping $\lambda$ finite, in the
limit the spacetime is infinitely flat $R_{0}\sim\lambda/k^{2}\rightarrow0$.
Individually, the Planck scale is not necessarily a characteristic
energy scale of this theory, neither the individual Hubble scale,
one can go beyond each individual scale and keeps their combination
(the only characteristic scale $\lambda$) the same. The gravity theory
is independent to how you define the Planck scale by an absolute ruler,
just like the fact that there is no specific scale such as the absolute
Planck scale in the Ricci flow. The new characteristic energy scale
being the critical density is very low, below which new phenomenon
may emerge. This is a major difference between this theory and the
traditional gravity theory. Since $R_{k}=R_{0}/\left(1-\frac{1}{4\pi}Gk^{2}\right)\gg0.34R_{0}$,
(\ref{eq:effective-action}) can always be approximately rewritten
as
\begin{equation}
S_{k}=\int d^{4}X\sqrt{\left|\det g\right|}\left[\mathcal{L}_{M}+2\lambda\frac{R_{k}}{R_{0}}\left(1-0.34\frac{R_{0}}{R_{k}}\right)\right]\approx\int d^{4}X\sqrt{\left|\det g\right|}\left[\mathcal{L}_{M}+2\lambda\frac{R_{k}}{R_{0}}\frac{1}{\sqrt{1+0.67\frac{R_{0}}{R_{k}}}}\right].\label{eq:relativity-MOND}
\end{equation}

These two effective actions have different limits at $R_{k}\rightarrow0$.
The gravitational part of the first action tends to a gravitational-source-independent
and finite constant at $R_{k}\rightarrow0$ so that it is considered
suitable for discussing global cosmological problems having a constant
background curvature. But the second effective action of the gravitational
part insteadly tends to zero at $R_{k}\rightarrow0$ and no CC in
it. Thus it is an approximation considered suitable for dealing with
local gravity systems which is governed by local gravitational source
($\mathcal{L}_{M}$) while CC is usually ignored. The approximation
of the second action is suitable for the scale of the local gravity
systems (for instance, galaxy), since at such scale where the local
curvature $R_{k}$, on one hand, can be considered not small so that
the background curvature of the universe in void affects the system,
on the other hand the local curvature $R_{k}$ can also be considered
small enough (e.g. at the edge of galaxy) so that there is a significant
modification of the standard gravity. Such scale is about the galactic
scale where CC is usually not considered but matter as gravitational
source is important. By using the weak field $g_{00}\approx1+2\Phi$,
$g_{ij}\approx-(1-2\Phi)\delta_{ij}$, $|\Phi|\ll1$ and static approximation
$\partial_{0}\Phi\approx0$, $\mathcal{L}_{M}\approx g_{00}\rho$,
we have the non-relativistic limit
\begin{equation}
S_{k}\approx\int d^{4}X\left[\frac{H_{0}^{2}}{4\pi G}\left(\frac{|\nabla\Phi|^{2}}{H_{0}^{2}}\frac{1}{\sqrt{1+2.01\frac{H_{0}^{2}}{|\nabla\Phi|^{2}}}}\right)+(1+2\Phi)\rho\right],
\end{equation}
where we have used $R_{k}\approx4|\nabla\Phi|^{2}$, $\lambda=\frac{R_{0}}{32\pi G}$
and $R_{0}=12H_{0}^{2}$. Surprisingly, the effective action of gravity
in certain sense is identical to a Modified Newtonian Dynamics (MOND)
theory giving rise to an explanation to the anomalous rotating curve
of galaxy without the dark matter (see recent review \cite{Milgrom:2014usa}
and references therein). When the acceleration $a_{N}=-\nabla\Phi$
of a rotating body in the Newtonian potential $\Phi$ is lower than
the scale $H_{0}$, then the Newton's law of gravity is strongly modified.
MOND works well in galactic scales, except the observation of collision
of two Bullet clusters \cite{Clowe:2006eq} which is often seem as
a failure of the non-relativistic MOND. However the full theory (\ref{eq:relativity-MOND})
is non-static and relativistic, and since the correction term $\left(1+0.67\frac{R_{0}}{R_{k}}\right)^{-1/2}$
in it plays the similar role of a polarized ``dielectric constant''
of gravity, so by analogy with electromagnetism, the full theory may
heuristically explain the phenomenon of Bullet clusters collision
as follows: the ``free charges'' (in analogy with visible matter)
collide while the ``polarized dielectric induced by the free charges''
(in analogy with dark matter halo that feels only gravity) pass through
each other. 

\subsection{Observable effect of Ricci flow limit: distance-redshift relation}

We have described that the reason why the spacetime metric continuously
deforms governed by the Ricci flow semi-classically is because the
existence of the non-trivial 2nd moment of the spacetime coordinates,
or equivalently, the quantum fluctuation of the spacetime. Here we
will see what is the physical effects of the 2nd moment of the spacetime
coordinates and the resulting flow of the spacetime metric in observations.
Reminding (\ref{eq:g+dg}) and (\ref{eq:Z},\ref{eq:definition tilde Z}),
we have the flow of the scale factor 
\begin{equation}
\langle\Delta a(\tau)\rangle^{2}=\langle\Delta a(0)\rangle^{2}+\langle\delta a^{2}\rangle=\tilde{Z}(\tau)\langle\Delta a(0)\rangle^{2}=\left(1+\delta_{\tilde{Z}}+\frac{2}{D}R\tau\right)\langle\Delta a(0)\rangle^{2},
\end{equation}
in which $\langle\Delta a(\tau)\rangle=\langle a(T,\tau)\rangle-\langle a(T_{0},\tau)\rangle$
is the classical displacement of the scale factors between different
physical time $T$ and $T_{0}$ which are both considered in relative
late epoches, and $\langle\delta a^{2}\rangle$ is the 2nd moment
contribution on it, so 
\begin{equation}
\frac{\langle\delta a^{2}\rangle}{\langle\Delta a(0)\rangle^{2}}=\delta_{\tilde{Z}}+\frac{2}{D}R\tau.
\end{equation}
An important observable in cosmology is the redshift, the measurement
of its mean value is given by scale factors at different epochs ($T$
and $T_{0}$) 
\begin{equation}
1+\langle z\rangle=\frac{\langle a(T_{0},0)\rangle}{\langle a(T,0)\rangle}.
\end{equation}
Its variance can be defined via Taylor expansion at a fixed epoch
$T$ but at different scale $\tau$ of the Ricci flow 
\begin{equation}
1+\frac{1}{2}\langle\delta z^{2}\rangle=\frac{\langle a^{2}(T,0)\rangle}{\langle a^{2}(T,\tau)\rangle},
\end{equation}
so we have
\begin{equation}
\frac{\langle\delta z^{2}\rangle}{\langle z\rangle^{2}}=-2\frac{\langle a^{2}(T,\tau)-a^{2}(T,0)\rangle}{\langle a(T,0)-a(T_{0},0)\rangle^{2}}=-2\frac{\langle\delta a^{2}\rangle}{\langle\Delta a(0)\rangle^{2}}=-2\delta_{\tilde{Z}}-\frac{4}{D}R\tau.\label{eq:dz^2/z^2}
\end{equation}

From the formula we see that the 2nd moment of the redshift renormalized
by the squared 1st moment redshift is monotone along $\tau$, which
is finite $-2\delta_{\tilde{Z}}\approx0.67$ at IR $\tau\rightarrow0$
and approaches to zero at UV $\tau\rightarrow\infty$. In other words,
at small scale the variance of the redshift can be ignored, but it
is significant (order $O(1)$) at cosmic scale observations. The proportional
relation between the 2nd and 1st moment of redshift gives a correction
to Distance-Redshift Relation at second order $O(z^{2})$ which can
not be ignored at large redshift, we expand the distance by the powers
of the redshift, 
\begin{equation}
\langle d_{L}(z)\rangle=\frac{1}{H_{0}}\left[\langle z\rangle+\frac{1}{2}\langle z^{2}\rangle+O(z^{3})\right]\overset{\tau\rightarrow0}{=}\frac{1}{H_{0}}\left[\langle z\rangle+\frac{1}{2}\left(1-2\delta_{\tilde{Z}}\right)\langle z\rangle^{2}+O(z^{3})\right],
\end{equation}
where we have used $\langle z^{2}\rangle=\langle z\rangle^{2}+\langle\delta z^{2}\rangle$
and $\langle d_{L}(z)\rangle$ is the distance between e.g. supernovas
and the earth observer in LWF. The 2nd moment of the redshift coming
from the 2nd moment of the spacetime coordinates does not modify the
relation at first order $O(z)$ which describes the expansion rate
of the universe, but modify its second order $O(z^{2})$ which describes
the accelerating or deceleration of the expansion. More precisely,
the 2nd moment of the redshift gives an additional deceleration parameter
$q_{0}=2\delta_{\tilde{Z}}\approx-0.67$, which is clearly redshift
independent and uniform behaving like a dark energy. 

The uniformness and universal of the quantum variance of the redshift
is also an indication that the Equivalence Principle (EP) could be
valid at the quantum level. The gravity is not only universally depicted
by the 1st moment of the metric (giving rise to the expansion rate)
but also the 2nd moment (giving rise to the acceleration). Phenomenologically
speaking, the spectral lines taking different energies universally
free-fall: not only they universally redshift (describing the expansion
rate) but also universally be broaden by the quantum variance (describing
the acceleration). 

We see that the spacetime coordinates become more and more fuzzy as
the Ricci flow driving the 2nd moment of the spacetime geometry becomes
more and more significant at cosmic scale. As a consequence, the quantum
variance of the redshift as physical observable becomes more and more
non-ignorable at large redshift regime. It is the reason why the universe
seems accelerating expansion. Indeed, we do not directly measure the
quantum variance of the redshift, instead of measuring the modified
Distance-Redshift Relation. So if this theory is true, it is an important
proposal to try to measure the almost linear dependence between the
quantum variance of redshift $\langle\delta z^{2}\rangle$ and the
squared-mean redshfit $\langle z\rangle^{2}$ shown in the IR $\tau\rightarrow0$
limit of (\ref{eq:dz^2/z^2}). At first glance, the measurement of
the variance or width $\langle\delta z^{2}\rangle$ may have numbers
of dirty non-quantum origins and effects, such as the thermo-broadening,
so that to single out the clean quantum part of the variance seems
difficult. But as the distance scale becomes larger and larger, the
ratio $\langle\delta z^{2}\rangle/\langle z\rangle^{2}$ becomes of
order one as predicted, so that the quantum part of the variance may
become dominant compared with other effects. On the other hands, unlike
other non-quantum effects, the quantum part of the variance is universal
as the EP claims which differs it from other noises. Therefore we
think the measurement of the quantum variance versus squared redshift
may be feasible. 

\section{Conclusions}

We summarize the results as follows. When the quantum fluctuations
are inescapable in the quantum measurement of the spacetime coordinates,
the Riemannian geometry can not be realized in rigor, so that the
2nd central moment or even higher moment of the spacetime coordinates
must not be ignored. We consider the effects and corrections of 2nd
central moment to the Riemannian geometry by a quantum non-linear
sigma models (Q-NLSM) interpreted as a quantum spacetime reference
frame (QSRF) system. The Ricci flow as the Gaussian approximation
of the renormalization flow of Q-NLSM, have been studied by powerful
tools, such as the reduced volume, which provides us a framework to
calculate how the geometry of the isotropic universe at current epoch
continuously deforms and finally how it looks like at very long distance
or cosmic scale. 

We show that as the spacetime with positive curvature isotropically
flows to IR by the Ricci flow, under the QSRF interpretation to the
Q-NLSM, (1) an effective Einstein-Hilbert-like action and a correct
cosmological constant (CC) emerge, (2) the universe becomes more and
more homogeneous and isotropic as the cosmological principle asserts,
and the metric tends to an Einstein metric at IR, moreover, (3) it
gives rise to an accelerating expansion universe with a fraction of
``dark energy'' $\Omega_{\Lambda}\approx0.67$ in the IR limit,
which is consistent with current observations. Therefore, in the sense
of effective quantum field theory of gravity, the CC problem is so
resolved. 

In the conceptual sense, the leading energy density coupled to gravity
is not anymore the quartic of the Planck scale cutoff, $\Lambda_{pl}^{4}$
coming from the zero-point fluctuation of the vacuum of the spacetime,
which is the main puzzle of the CC problem. It is resolved by noticing
in the theory that, the parameter background $x,y,z,t$, which is
absolute, external, classical and free from any quantum fluctuation,
in fact can not be interpreted as physical spacetime in rigor. The
unphysical nature of the parameter background makes the zero-point
energy densities $\Lambda_{pl}^{4}$ completely unobservable, including
the Casimir effect \cite{PhysRevD.72.021301}, and hence disappears
in the effective action (\ref{eq:effective action}). In contrast,
the leading energy density coupled to gravity is given by the two-point
quantum fluctuation of the physical spacetime $\langle\delta X_{\mu}\delta X_{\nu}\rangle\neq0$
or $\delta g_{\mu\nu}^{(2)}\neq0$ depicted semi-classically by the
Ricci flow. In this sense, the Equivalence Principle (EP) is retained
and generalized to the quantum level: energy densities which universally
coupled to gravity are those with respect to the physical spacetime
$X^{\mu}$ which is inescapably quantum fluctuating.

A measurement to test the theory is also proposed. In this theory,
the phenomenological existence of CC or the ``dark energy'' is all
about the deviation of the Distance-Redshift relation from the standard
Hubble's law at relative large redshift regime. The theory suggests
that the deviation is due to the quantum variance of the redshift
$\langle\delta z^{2}\rangle$ induced by the 2nd central moment of
the spacetime coordinates. An almost linear dependence between the
quantum variance $\langle\delta z^{2}\rangle$ and the squared-redshift
$\langle z\rangle^{2}$ is predicted (\ref{eq:dz^2/z^2}), and the
proportionality constant at long distance limit is $-2\delta_{\tilde{Z}}\approx0.67$
being close to the ``acceleration'' parameter $-q_{0}$. And we
argue that to measure the clean quantum part of the variance of the
redshift seems feasible.

Here we discuss the prospect of other Ricci flow's applications to
the cosmology, for instance, the very early universe. The CC is shown
as a special example of the application of the Ricci flow with isotropic
and positive curvature initial condition, where the space and time
coordinates are renormalized on an equal footing. Such situation is
relatively easy to deal with, since a CC-normalized Ricci flow with
isotropic positive curvature initial condition is free from singularities
and diffusion term dominants. It is not necessarily the case when
the universe is in the very early epoch, where the spatial part of
the universe approaches to a singularity. The space and time differ
from each other in the very early universe, in such anisotropic initial
condition, the Ricci flow may develop local singularities. The application
of the Ricci flow (or its generalization) to a more general initial
condition, such as the very early universe, is still a challenge.
Because, firstly in the regime near the singularity, the validity
of the Ricci flow approximation is unclear; and secondly the mathematical
tools to deal with the singularities developed by the Ricci flow is
also highly technical. Some qualitative results from the studies of
such case gives us evidences that the application of Ricci flow to
the very early universe is also worth pursuing: (1) the local singular
scale factor $a(\mathcal{T}\rightarrow0,\tau\rightarrow0)$ in (\ref{eq:FRW})
with the corresponding scalar function $f(\mathcal{T}\rightarrow0,\tau\rightarrow0)\rightarrow\mathrm{const}$
at very early epoch developed by the Ricci flow can be well modeled
by a pinched soliton solution $a(\mathcal{T}\rightarrow0,\tau\rightarrow0)\sim e^{H_{*}\mathcal{T}}$
with an expanding rate $H_{*}\sim\frac{1}{\sqrt{\tau_{*}}}$ whose
profile resembles a spatially inflationary (de Sitter) universe; (2)
the linear fluctuations at the vicinity of the singularity is governed
by a linearlized version of the Ricci soliton equation, the Lichnerowicz
equation, which gives rise to an evolution equation for the fluctuation
being similar with the primordial fluctuation of the universe; (3)
a small deviation of $\tau_{*}$ from $\tau=0$ (IR limit) not only
at leading order gives a de Sitter spacetime with a large expanding
rate, but also at next order gives small deviation of $a(\mathcal{T}\rightarrow0,\tau_{*})$
from the exact de Sitter, which gives rise to a small slow roll parameter
and small deviation of the fluctuation spectrum from exact scale invariant.
The detail studies are leaving for our future works.
\begin{acknowledgments}
This work was supported in part by the National Science Foundation
of China (NSFC) under Grant No.11205149, and the Scientific Research
Foundation of Jiangsu University for Young Scholars under Grant No.15JDG153.\bibliographystyle{plain}

\end{acknowledgments}

\end{document}